\documentclass[useAMS,usenatbib,usegraphicx]{mn2e}

\title[Radio jet proper motion at $z$$>$5]{The first estimate of radio jet proper motion at $z$$>$5}

\author[S. Frey et al.]{S\'andor Frey$^{1}$\thanks{E-mail:
frey@sgo.fomi.hu}, Zsolt Paragi$^{2}$, Judit O. Fogasy$^{3}$ and Leonid I. Gurvits$^{2,4}$\\
$^{1}$F\"OMI Satellite Geodetic Observatory, P.O. Box 585, H-1592 Budapest, Hungary\\
$^{2}$Joint Institute for VLBI in Europe, Postbus 2, 7990 AA Dwingeloo, the Netherlands\\
$^{3}$Department of Earth and Space Sciences, Chalmers University of Technology, Onsala Space Observatory, SE-43992 Onsala, Sweden\\
$^{4}$Department of Astrodynamics and Space Missions, Delft University of Technology, 2629 HS Delft, the Netherlands}

\begin{document}

\date{Accepted 2014 October 23. Received 2014 October 21; in original form 2014 July 6}

\pagerange{\pageref{firstpage}--\pageref{lastpage}} \pubyear{2014}

\maketitle

\label{firstpage}

\begin{abstract}
The extremely high redshift ($z$=5.3) radio source SDSS~J102623.61+254259.5 (J1026+2542) is among the most distant and most luminous radio-loud active galactic nuclei (AGN) known to date. Its one-sided radio jet structure on milli-arcsecond (mas) and $\sim$10-mas scales typical for blazars was first imaged at 5~GHz with very long baseline interferometry (VLBI) in 2006. Here we report on our dual-frequency (1.7 and 5~GHz) imaging observations performed with the European VLBI Network (EVN) in 2013. The prominent jet structure allows us to identify individual components whose apparent displacement can be detected over the time span of 7.33~yr. This is the first time when jet proper motions are directly derived in a blazar at $z$$>$5. The small values of up to $\sim$0.1~mas\,yr$^{-1}$ are consistent with what is expected in a relativistic cosmological model if redshift is a measure of distance. The apparent superluminal jet speeds, considered tentative because derived from two epochs only, exceed 10\,$c$ for three different components along the jet. Based on modeling its spectral energy distribution, J1026+2542 is known to have its X-ray jet oriented close to the line of sight, with significant Doppler boosting and a large bulk Lorentz factor ($\Gamma$$\approx$13). The new VLBI observations, indicating $\sim 2.3 \times 10^{12}$~K lower limit to the core brightness temperature, are consistent with this picture. The spectral index in the core region is $-0.35$.
\end{abstract}


\section{Introduction}

Blazars are radio-emitting active galactic nuclei (AGN) prominent in other wavebands as well. The class is divided into BL Lacertae (BL Lac) objects and flat-spectrum radio quasars (FSRQ). Their relativistic jets originating from the vicinity of a central supermassive black hole are known to be oriented close to the line of sight \citep[e.g.][]{Urry95}. The number of known blazars is steadily increasing, owing to the sky surveys at radio, infrared, optical, X-ray, and $\gamma$-ray bands \citep[e.g.][and references therein]{Giom12}. However, confirmed blazars at extremely high redshifts ($z$$>$5) are still rare. The most distant one is J0906+6930 at $z$=5.47 \citep{Roma04}, the second one is J1026+2542 at $z$=5.266 \citep{Sbar12,Sbar13}. Based on X-ray and other data, \citet{Ghis14} recently suggested that J1146+4037 ($z$=5.01) is also a blazar.

The radio emission from the blazar jets can be imaged with the highest angular resolution using very long baseline interferometry (VLBI). Typical observed features are compact flat-spectrum ``cores'', one-sided jets, high Doppler-boosted brightness temperature, and often apparent superluminal motion of jet components. These can be reconciled with a synchrotron jet model where the relativistic outflow of plasma is inclined at a very small angle to the line of sight \citep[e.g.][]{Blan79,Urry95}. All the three $z$$>$5 blazars known to date, J0906+6930 \citep{Roma04}, J1026+2542 \citep{Helm07,Frey13}, and J1146+4037 \citep{Frey10} have already been imaged with VLBI. The source J1026+2542 stands out from this small sample because it shows a prominent jet structure with multiple components, extending out to at least 20 milli-arcseconds (mas). Its first VLBI imaging observations at 5~GHz date back to 2006 \citep{Helm07}. The rich jet structure and the first-epoch reference measurement give us a unique opportunity to look for possible changes in the radio jet of J1026+2542, by means of repeated VLBI imaging at the same observing frequency. 

VLBI monitoring observations are a useful direct tool to study kinematic properties of AGN jets. By measuring the apperent component speeds and estimating the Doppler-boosting factor in an independent way, it is possible to infer the Lorentz factor of the plasma flow and the jet orientation with respect to the line of sight. Homogeneous multi-epoch surveys of large radio-loud AGN samples \citep[e.g.][]{Kell04,Pine07,Brit08,List09,Mars11,Pine12,List13} could reveal complex kinematics in individual sources, like jet curvature, acceleration and deceleration, non-radial motion, stationary or inward-moving features. Apart from studying the jet physics, the measured apparent proper motions in a large sample of jet components over a broad range of redshifts can also serve as a test for cosmological models \citep[e.g.][]{Cohe88,Verm94,Kell99}. It appears that low values of jet component proper motions are observed at all redshifts, but high values ($\ga0.2$~mas\,yr$^{-1}$) are at low redshifts ($z\la2$) only. This is consistent with the cosmological interpretation of AGN redshifts \citep[e.g.][]{Kell04}. Notably, the redshift ranges of bright flux density-limited AGN samples intensively monitored for jet kinematic studies \citep[e.g.][]{Kell04,Brit08} do not reach beyond $z$$\sim$3.5. Therefore new kinematic data for sources at extremely high redshifts extend the apparent proper motion--redshift relation to an unexplored terrain.     

The subject of our paper, SDSS~J102623.61+254259.5 (J1026+2542 in short) is a radio-loud AGN at $z$=5.266 \citep{Ahn12}. It was identified as the second most distant blazar by \citet{Sbar12}, based on its spectral energy distribution (SED) and its X-ray spectrum. \citet{Sbar12} estimated the mass of the central black hole in J1026+2542 as $(2-5) \times 10^9$~$M_{\odot}$, and suggested that the bulk Lorentz factor of the jet is $\Gamma$=14. In a recent paper, based on a more complete SED supplemented with new high-energy X-ray observations made by the {\it Nuclear Spectroscopic Telescope Array (NuSTAR)} satellite, \citet{Sbar13} reinforced the parameters derived for the jet in J1026+2542, giving $\Gamma$$\approx$13 and the viewig angle to the line of sight $\theta$=$3\degr$. In our earlier study \citep{Frey13}, we analysed the archival VLBI imaging data of \citet{Helm07} and found that the prominent one-sided jet of J1026+2542 is typical for a blazar. However, the brightness temperature estimated from fitting a brightness distribution model to the seemingly extended radio ``core'' did not provide an unequivocal evidence for a strong Doppler boosting. Repeated VLBI imaging with somewhat higher angular resolution promised a chance to distinguish between the possible models and thus to estimate the radio jet parameters more reliably.  

Here we report on our new VLBI observations of the high-redshift blazar J1026+2542. The observations were performed with the European VLBI Network (EVN) at 5~GHz on 2013 May 28, and at 1.7~GHz on 2013 June 4. At 5~GHz, the second-epoch imaging gives a time baseline of 2677 days (7.33~yr) with respect to the earlier observations of \citet{Helm07} made with the US Very Long Baseline Array (VLBA) on 2006 January 28. In the rest frame of the source at $z$=5.266, due to the time dilation caused by the expansion of the Universe, this corresponds to a period shorter by a factor of (1+$z$), i.e. 1.17~yr. Throughout the paper, we assume a flat cosmological model with $H_{\rm 0}$=70~km~s$^{-1}$~Mpc$^{-1}$, $\Omega_{\rm m}$=0.3, and $\Omega_{\Lambda}=$0.7. In this model, $1\arcsec$ angular size corresponds to 6.12~kpc projected linear size at $z$=5.266, the luminosity distance of the source is $D_{\rm L}$=49.559~Gpc, and the age of the Universe is 1.08~Gyr \citep{Wrig06}.

\section{Observations and data reduction}

The 5-GHz EVN observation of the blazar J1026+2542 took place on 2013 May 28, using a global network of fourteen radio telescopes including Effelsberg (Germany), Jodrell Bank Lovell Telescope (UK), Medicina, Noto (Italy), Onsala (Sweden), Toru\'n (Poland), Yebes (Spain), Badary, Svetloe, Zelechukskaya (Russia), Hartebeesthoek (South Africa), Sheshan, Urumqi (China), and the Westerbork Synthesis Radio Telescope (WSRT, the Netherlands). The experiment EF024A lasted for 4 h and used Mark5A disk recording at 1024~Mbit~s$^{-1}$ data rate with two circular polarizations, eight basebands per polarization, each with thirty-two 500-kHz spectral channels. This resulted in a total bandwidth of 128~MHz in both left and right circular polarizations, using 2-bit sampling. The correlation of the data with 2-s integration time was performed at the EVN software correlator \citep[SFXC,][]{Pido09} at the Joint Institute for VLBI in Europe (JIVE) in Dwingeloo, the Netherlands. The 1.7-GHz experiment (project code: EF024B) was conducted nearly contemporaneously in the same EVN observing session, on 2013 June 4. With a duration and setup similar to that of EF024A, the participating VLBI stations were the same as above, except for Yebes. The correlator integration time was 4 s in the case of the 1.7-GHz experiment. Apart from the target, J1026+2542, the fringe-finder sources J0854+2006 and J1159+2914 were also observed occasionally. The total observing time spent on J1026+2542 was 3.3~h in each experiment.

The NRAO Astronomical Image Processing System\footnote{\tt{http://www.aips.nrao.edu}} ({\sc AIPS}) was used for the data calibration in a standard way \citep[e.g.][]{Diam95}. The visibility amplitudes were calibrated using antenna gains and system temperatures measured at most of the telescopes. The exceptions were Badary, Svetloe, Zelechukskaya (in both experiments), and Jodrell Bank (at 1.7~GHz), where only nominal system temperature values were available. Fringe-fitting \citep{Schw83} was first performed for the bright fringe-finder sources (J0854+2006, J1159+2914) using 3-min solution intervals. These data were exported to the Caltech {\sc Difmap} package \citep{Shep94} for imaging. The conventional hybrid mapping procedure involving several iterations of {\sc CLEAN}ing \citep{Hogb74} and phase (then amplitude) self-calibration resulted in the images and brightness distribution models for the calibrators. Overall antenna gain correction factors (up to $\sim$20 per cent but typically less) were determined for J0854+2006 and J1159+2914 in {\sc Difmap} and applied to the visibility amplitudes in {\sc AIPS}. Fringe-fitting was then performed for J1026+2542 as well, using 5-min solution intervals. The fully calibrated data sets were averaged in frequency and transferred to {\sc Difmap} for imaging and model fitting. Thanks to the gain corrections based on the fringe-finder data, the antenna gains for J1026+2542 became consistent well within 5 per cent. We will assume that the absolute flux density calibration uncertainty in our experiments is 5 per cent \citep[like e.g. in][]{An12}.

\section{Results}

\subsection{Images and models}

\begin{figure}
\centering
  \includegraphics[bb=70 75 524 720, width=60mm, angle=270, clip=]{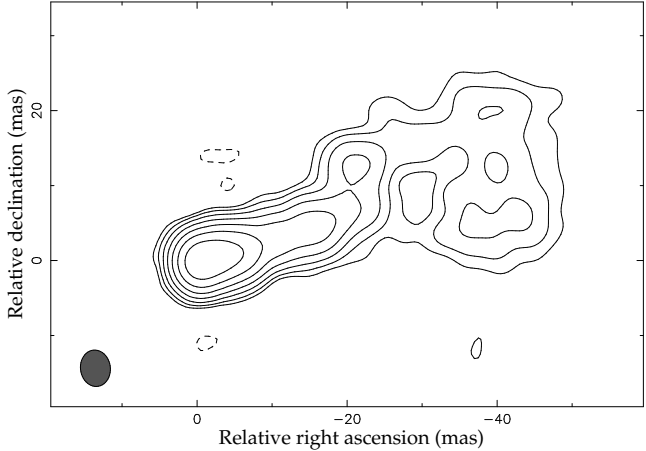}
  \caption{
The naturally-weighted 1.7-GHz {\sc CLEAN}-component EVN image of J1026+2542. The lowest contours are drawn at $\pm0.45$~mJy~beam$^{-1}$ ($\sim$3$\sigma$ image noise level). Further positive contour levels increase by a factor of 2. The peak brightness is 57.4~mJy~beam$^{-1}$. The Gaussian restoring beam is 4.87~mas $\times$ 3.98~mas with major axis position angle PA=$7\fdg2$. It is indicated with a filled ellipse (full width at half maximum, FWHM) in the lower-left corner.}
  \label{Lband-image}
\end{figure}

\begin{figure}
\centering
  \includegraphics[bb=70 75 524 720, width=60mm, angle=270, clip=]{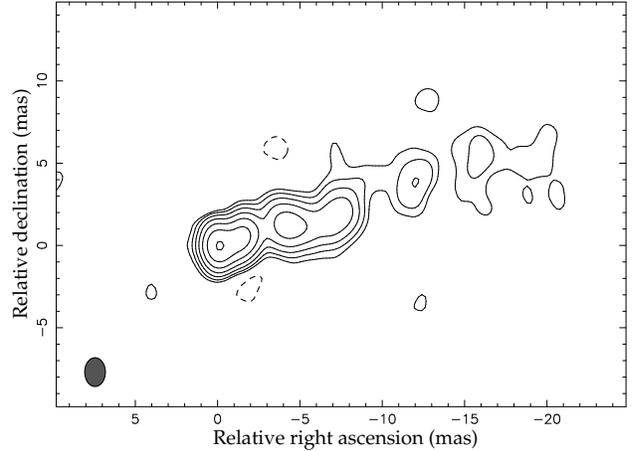}
  \caption{
The naturally-weighted 5-GHz {\sc CLEAN}-component EVN image of J1026+2542. The lowest contours are drawn at $\pm0.32$~mJy~beam$^{-1}$ ($\sim$3$\sigma$ image noise level). Further positive contour levels increase by a factor of 2. The peak brightness is 21.7~mJy~beam$^{-1}$. The Gaussian restoring beam is 1.74~mas $\times$ 1.25~mas with PA=$0\fdg1$.}
  \label{Cband-image}
\end{figure}

The total intensity images resulted from hybrid mapping, i.e. several iterations of {\sc CLEAN}ing and self-calibration, are shown in Fig.~\ref{Lband-image} (1.7~GHz) and Fig.~\ref{Cband-image} (5~GHz). The absolute coordinates of the source are $10^{\rm h} 26^{\rm m} 23\fs62113$ right ascension and $25\degr 42\arcmin 59\farcs4291$ declination\footnote{\tt{http://astrogeo.org/vlbi/solutions/rfc\_2014c}}. The characteristic position angle (around $-70\degr$) and the angular extent ($\sim$20~mas) of the one-sided core--jet structure in the 5-GHz image are qualitatively consistent with what is seen in an earlier 5-GHz VLBI image \citep{Helm07,Frey13}. At the lower frequency, the jet can be traced out to $\sim$50~mas (Fig.~\ref{Lband-image}), corresponding to $\sim$300~pc projected linear size. The reason for the apparent longer extension of the 1.7-GHz
jet is the steep radio spectrum of the optically thin emission. The structure of the 1.7-GHz jet becomes diffuse and the opening angle increases with the distance from the core.

To parametrize the brightness distribution of J1026+2542 with a set of simple model components, we used the self-calibrated visibility data for model fitting in {\sc Difmap}. At 1.7~GHz, the complex structure is fitted with a point-like central component and six circular Gaussian model components (Table~\ref{Lband-modelfit}). The model brightness distribution obtained this way \citep{Pear95} is not unique, but provides a good representation of the source structure that can be used for estimating physical parameters and quantifying structural changes in the source if models obtained at different epochs are available. The size upper limit of the point-like component is estimated by calculating the minimum resolvable angular size in this VLBI experiment \citep{Kova05}. An attempted circular Gaussian fit to the central component converged to a size below the minimum resolvable size, justifying the choice of a point-source model and indicating the compactness of the core.

\begin{table}
  \centering 
  \caption[]{Parameters of the model components fitted to the 1.7-GHz VLBI visibility data of J1026+2542. The statistical errors are estimated according to \citet{Foma99}, with additional 5 per cent flux density calibration uncertainties assumed.}
  \label{Lband-modelfit}
\begin{tabular}{cccc}        
\hline                 
Flux density & \multicolumn{2}{c}{Relative position}& Diameter    \\
$S$ (mJy)    & R.A. (mas)        & Dec. (mas)       & FWHM (mas)  \\
\hline                       
39.1$\pm$2.0 &  ...              &  ...             &  $<$0.24$\times$0.20  \\  
39.4$\pm$2.0 &  $-$2.59$\pm$0.01 &  0.87$\pm$0.01   &  1.58$\pm$0.01  \\
26.2$\pm$1.3 &  $-$5.79$\pm$0.01 &  1.37$\pm$0.01   &  1.88$\pm$0.01  \\
10.4$\pm$0.6 &  $-$9.92$\pm$0.03 &  3.18$\pm$0.03   &  2.79$\pm$0.05  \\
22.4$\pm$1.2 & $-$15.03$\pm$0.06 &  3.85$\pm$0.06   &  5.53$\pm$0.12  \\
13.1$\pm$0.9 & $-$20.53$\pm$0.13 & 10.56$\pm$0.13   &  6.00$\pm$0.25  \\
29.8$\pm$3.5 & $-$38.04$\pm$0.84 & 11.71$\pm$0.84   & 15.88$\pm$1.67  \\
\hline   
\end{tabular}
\end{table}

The model at 5~GHz (Table~\ref{Cband-modelfit}, bottom section) contains a point-like central component and five circular Gaussians. During the model fitting, we kept the number of parameters at the minimum. Here we also attempted to fit the central component with a Gaussian but its size converged to zero. We then performed model fitting for the 5-GHz VLBA data obtained by \citet{Helm07} in 2006 in a consistent way, i.e. with a point source in the centre, to facilitate the direct comparison between the two epochs. The archival calibrated visibility data are publicly available from the web site of the VLBA Imaging and Polarimetry Survey (VIPS)\footnote{\tt{http://www.phys.unm.edu/{\textasciitilde}gbtaylor/VIPS}}. The parameters of the brightness distribution model fitted to the VLBA data (a point source and five circular Gaussians) are also given in Table~\ref{Cband-modelfit} (top section). The component positions are measured with respect to the central point source at both epochs. Note that the relative positions cited here differ from those reported in \citet{Frey13}, where the two innermost components in the top section of Table~\ref{Cband-modelfit} were instead fitted with a single elliptical Gaussian. However, the flux densities and sizes of the other circular Gaussians in this paper are consistent with those of \citet{Frey13}. 

The statistical errors of the fitted Gaussian model component positions (Tables~\ref{Lband-modelfit} and \ref{Cband-modelfit}) reflect the contribution of the image noise to the uncertainties \citep{Foma99}. However, realistic errors that account for the imperfect coverage of the interferometer array, and the complexity of the jet structure (i.e. the possible blending of nearby components) are more difficult to estimate. We follow a conservative approach and adopt as much as 10 per cent of the restoring beam size \citep{List09} as the uncertainty of the component positions wherever it exceeds the statistical error determined for a given component. In particular, for the 5-GHz EVN data, this leads to the uncertainties of 0.13~mas and 0.17~mas in the relative right ascensions and declinations, respectively. For the 5-GHz VLBA data, these values are 0.18~mas and 0.29~mas, respectively.

\begin{figure}
\centering
  \includegraphics[bb=70 75 524 720, width=60mm, angle=270, clip=]{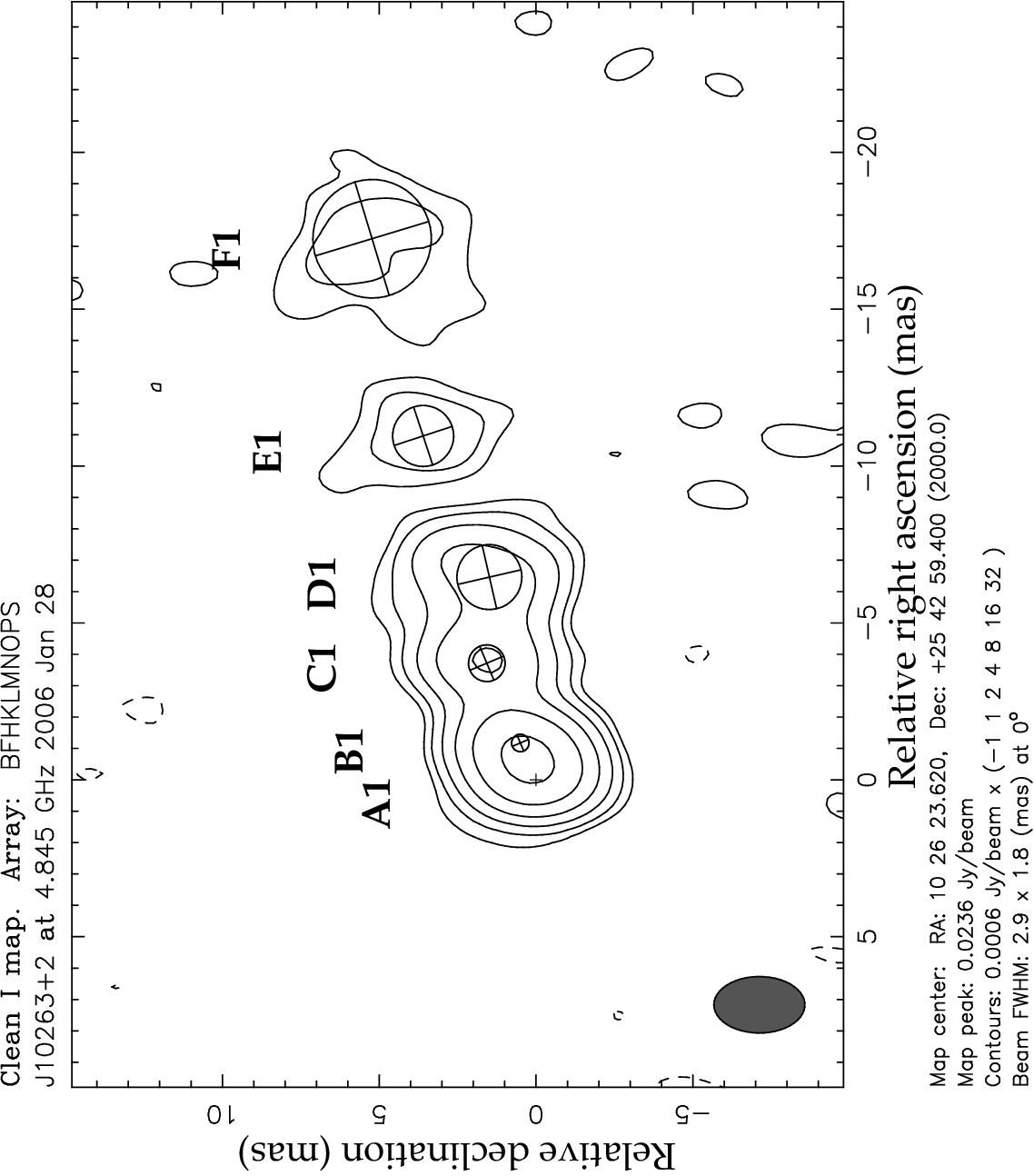}
  \includegraphics[bb=70 75 524 720, width=60mm, angle=270, clip=]{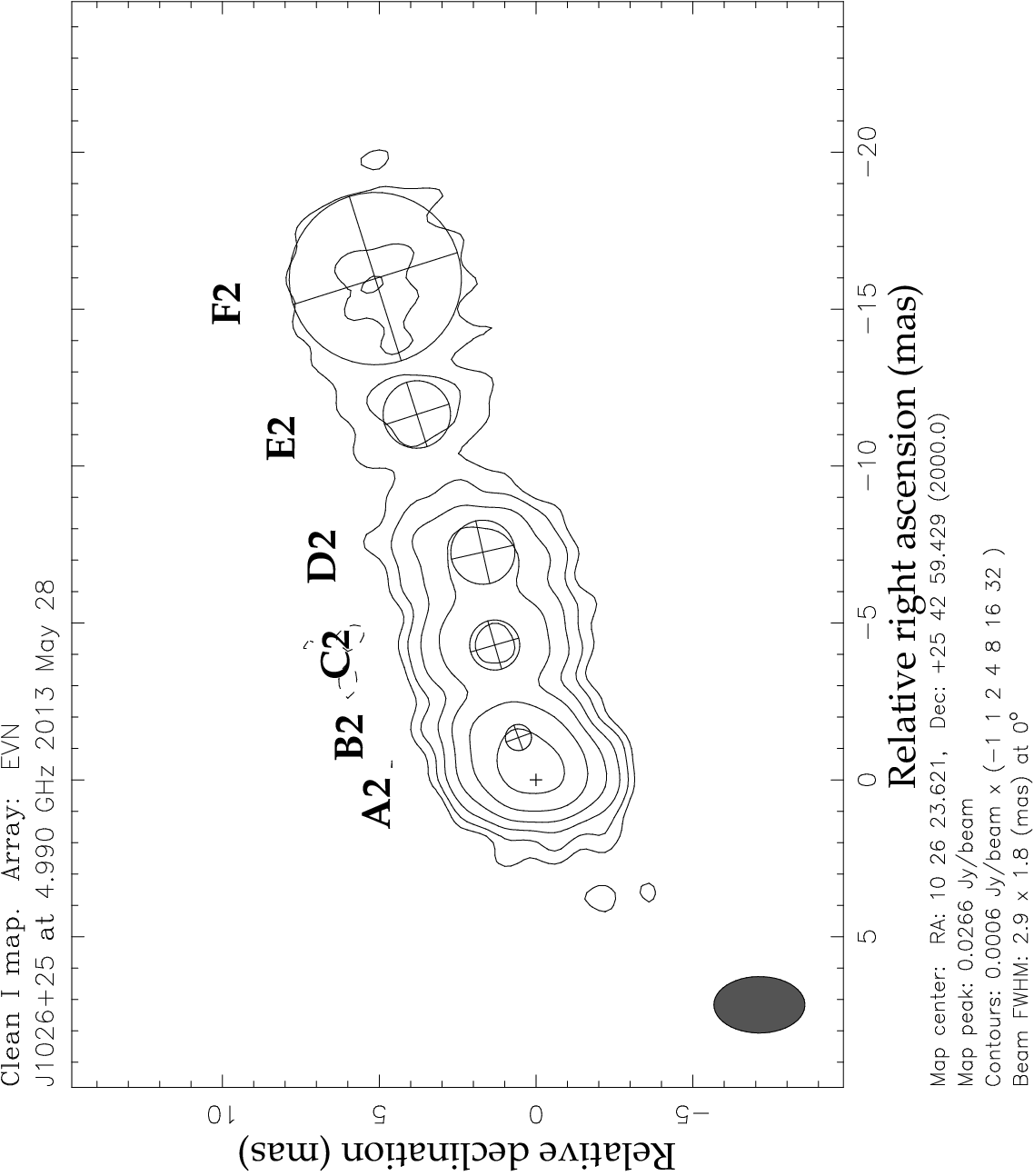}
  \caption{
5-GHz VLBI images of J1026+2542 restored with the point and circular Gaussian model components listed in Table~\ref{Cband-modelfit}. The observations were made at two epochs separated by 7.33~yr. In both images, the lowest contours are at $\pm0.6$~mJy~beam$^{-1}$, the positive contour levels increase by a factor of 2, and the restoring beam is 2.9~mas $\times$ 1.8~mas at PA=$0\degr$. The peak brightness is 23.6~mJy~beam$^{-1}$ (epoch 2006 January 28, top), and 26.6~mJy~beam$^{-1}$ (epoch 2013 May 28, bottom). Point components (A1 and A2) are plotted with crosses and located at the origin. The location and the diameter (FWHM) of the individual Gaussian model components (B1 to F1 and B2 to F2) are indicated with circles. The axes are parallel and perpendicular to the given component's position angle with respect to the central point source.
} 
  \label{VLBI-models}
\end{figure}

By means of comparing the models presented here, we are able to detect changes in the source structure, to identify individual jet components, and to estimate their apparent proper motions over the 7.33-yr time span in the observer's frame. To aid visual comparison, the 5-GHz VLBI images of J1026+2542, restored with the respective models from 2006 and 2013 (Table~\ref{Cband-modelfit}), are displayed together in Fig.~\ref{VLBI-models}. For consistency, we used the same scaling and the same restoring beam size (adjusted to the VLBA case with somewhat lower resolution) in these two images.

\begin{table}
  \centering 
  \caption[]{Parameters of the model components fitted to 5-GHz VLBI visibility data of J1026+2542 obtained by \citet{Helm07} in 2006 (top) and in our experiment in 2013 (bottom). The statistical errors are estimated according to \citet{Foma99}, with additional 5 per cent flux density calibration uncertainties assumed.}
  \label{Cband-modelfit}
\begin{tabular}{ccccc}        
\hline                 
ID & Flux density & \multicolumn{2}{c}{Relative position}& Diameter    \\
   & $S$ (mJy)    & R.A. (mas)        & Dec. (mas)       &  (mas)  \\
\hline                       
A1 & 14.9$\pm$0.8 &  ...              &  ...             &  $<$0.29$\times$0.18 \\  
B1 & 18.1$\pm$1.0 &  $-$1.17$\pm$0.01 &  0.50$\pm$0.01   &  0.55$\pm$0.01  \\
C1 & 12.2$\pm$0.7 &  $-$3.72$\pm$0.01 &  1.57$\pm$0.01   &  1.17$\pm$0.01  \\
D1 & 13.3$\pm$0.9 &  $-$6.46$\pm$0.03 &  1.49$\pm$0.03   &  2.07$\pm$0.05  \\
E1 &  3.8$\pm$0.6 & $-$10.96$\pm$0.12 &  3.60$\pm$0.12   &  1.94$\pm$0.24  \\
F1 &  7.0$\pm$1.2 & $-$17.24$\pm$0.29 &  5.23$\pm$0.29   &  3.78$\pm$0.58  \\
\hline   
A2 & 20.0$\pm$1.0 &  ...           &  ...                &  $<$0.12$\times$0.09 \\  
B2 & 20.0$\pm$1.0 &  $-$1.35$\pm$0.01 &  0.56$\pm$0.01   &  0.83$\pm$0.01  \\
C2 & 16.0$\pm$0.9 &  $-$4.30$\pm$0.01 &  1.32$\pm$0.01   &  1.60$\pm$0.01  \\
D2 & 11.4$\pm$0.7 &  $-$7.25$\pm$0.02 &  1.70$\pm$0.02   &  2.05$\pm$0.03  \\
E2 &  2.8$\pm$0.4 & $-$11.68$\pm$0.14 &  3.74$\pm$0.14   &  2.15$\pm$0.29  \\
F2 &  9.0$\pm$2.2 & $-$16.01$\pm$0.66 &  5.06$\pm$0.66   &  5.49$\pm$1.32  \\

\hline   
\end{tabular}
\end{table}

\subsection{Jet component proper motions}

From the 5-GHz images presented in this paper (Fig.~\ref{Cband-image}), and by \citet{Helm07} and \citet{Frey13}, it is obvious that there are detectable changes in the jet structure between the two epochs. The fitted model parameters (Table~\ref{Cband-modelfit} and Fig.~\ref{VLBI-models}) can be used to estimate the apparent proper motion of jet components identified at both epochs. 
We assume that the base of the jet (the point-like component that serves as the reference point) is stationary. Here we concentrate on components B, C, D, and E. The position and size of the low-brightness, most extended component F at the diffuse end of the VLBI-imaged jet are rather uncertain. Formally, it shows an inward motion between the two epochs, but it is quite consistent with a stationary feature considering the large errors (Table~\ref{Cband-modelfit}). It could indicate a recollimation shock \citep[e.g.][]{Peru13}, but the current kinematic and spectral data are insufficient to elaborate on this possibility. 

In any case, the proper motions derived from two-epoch measurements should be treated with caution. Long-term studies of individual objects \citep[recent references include e.g.][]{Agud12,Schi12,Qian13,Wu13}, and extensive multi-epoch surveys like the MOJAVE (Monitoring Of Jets in Active galactic nuclei with VLBA Experiment) program \citep[e.g.][]{List09,List13}, the Radio Reference Frame Image Database \citep[RRFID; e.g.][]{Pine07,Pine12} or the CJF (Caltech--Jodrell Bank Flat-spectrum) survey \citep[e.g.][]{Brit08} provide more reliable proper motions for large samples of sources which even allow for investigations of e.g. jet accelerations, transverse motions, and orientation variations. However, these detailed studies are usually restricted to the brightest radio-loud AGN, with redshifts rarely exceeding 3. The case of J1026+2542 is unique because no direct VLBI jet proper motion estimate is available so far for any radio source at $z$$>$5.       

With the caveat above, the positional changes of components B, C, D, and E with respect to the central component correspond to the apparent proper motions listed in Table~\ref{properm}. The errors are based on the conservative estimates of the component positional uncertainties (i.e. 10 per cent of the restoring beam size) at the two epochs. For components C, D, and E, the apparent transverse speeds ($\beta_{\rm app} \ga 10 \,c$) are significant and consistent with each other. The transverse speed of the jet component B is poorly defined, likely due to its blending with the core.

\begin{table}
  \centering 
  \caption[]{Apparent proper motions and transverse speeds of the jet components in J1026+2542.}
  \label{properm}
\begin{tabular}{ccc}        
\hline                 
Component ID & Proper motion  & Transverse speed   \\
   & mas\,yr$^{-1}$ & $c$                \\
\hline                       
B & 0.026$\pm$0.031 &   3.3$\pm$3.9     \\  
C & 0.086$\pm$0.033 &  10.8$\pm$4.1    \\
D & 0.112$\pm$0.031 &  14.0$\pm$3.9    \\
E & 0.100$\pm$0.031 &  12.5$\pm$3.9    \\

\hline   
\end{tabular}
\end{table}

\subsection{Flux densities and spectral index}

From the flux density of the unresolved core at $\nu_1$=1.7~GHz ($S_{1.7}$=39.1~mJy; Table~\ref{Lband-modelfit}) and at $\nu_2$=5~GHz ($S_{5}$=20.0~mJy; Table~\ref{Cband-modelfit}), the two-point spectral index is $\alpha_{\rm core}$=$-0.6$ (the spectral index is defined as $S\propto\nu^{\alpha}$). Flux density variability is likely negligible here since the EVN observations at the two frequencies were made within a week. 

\begin{figure}
\centering
  \includegraphics[bb=77 75 530 715, width=60mm, angle=270, clip=0]{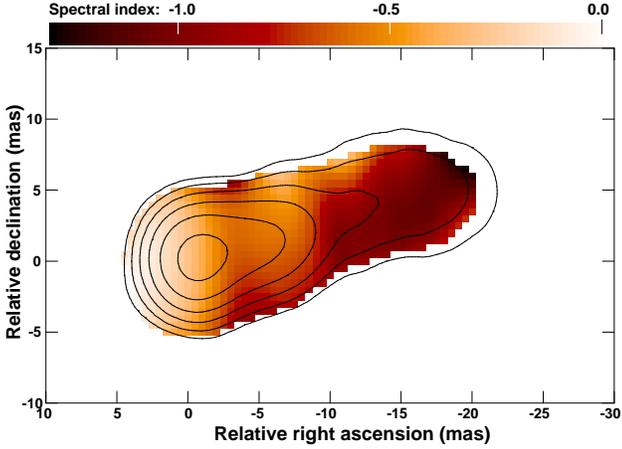}
  \caption{
The spectral index image of J1026+2542, constructed from the nearly contemporaneous 1.7-GHz and 5-GHz EVN images. The restoring beam at 5~GHz was set to match the more extended 1.7-GHz beam (see Fig.~\ref{Lband-image}). The total intensity contours illustrate the source structure at 5~GHz. The contours start at 0.9~mJy~beam$^{-1}$, the levels increase by a factor of 2, the peak brightness is 38.5~mJy~beam$^{-1}$. The two-point spectral indices displayed in the colour scale are calculated for pixels brighter than the 12$\sigma$ noise level in both images.
} 
  \label{sp-index}
\end{figure}

Based on the 1.7-GHz and 5-GHz EVN data, we present the spectral index image of J1026+2542 in Fig.~\ref{sp-index}. This allows us to analyse the spectral index distribution along the jet in J1026+2542. To prepare the spectral index map, first the restoring beam and the pixel size at 5~GHz were matched to those of the 1.7-GHz image (Fig.~\ref{Lband-image}). Then the 5-GHz image was slightly shifted to account for the frequency-dependent core shift. The apparent change in the core position in flat-spectrum radio quasars is due to the frequency-dependent opacity in synchrotron self-absorbed compact jets \citep[e.g.][]{Loba98}. As shown by \citet{Soko11} on the basis of 9-frequency analysis of VLBI-scale structures in 20 AGN between the observing frequencies of 1.6 and 15~GHz, the core shift effect is considerably more prominent at the lower frequencies. In their study, the average value of the core displacement between 1.6 and 5~GHz was 0.88 mas. The displacement shows a systematic direction along the jet: in all sources, 1.6-GHz cores are found to be closer to the outermost optically thin components than their 5-GHz counterparts. The sources studied by \citet{Soko11} are distributed within the redshift range from 0.056 to 2.707. Given a clear trend of decrease of the core shift with the increasing frequency \citep[Fig.~3 in][]{Soko11}, and the high rest-frame frequencies in our $z$=5.266 source J1026+2542, the expected core shift is significantly smaller than 1~mas. In the rest frame of the source, our observed frequencies correspond to $\sim$10~GHz and 31~GHz.

The images could in principle be aligned using the positions of optically thin jet components seen at both frequencies. In our case, this proved uncertain since the jet has a complex structure with several components that are difficult to cross-identify in the two images with sufficient positional accuracy. A more sophisticated method developed by \citet{Crok08} would maximize a cross-correlation function between the two images of the entire optically thin region of the jet to estimate a core shift. Here we applied a somewhat similar but simpler method. We transferred the circular Gaussian model components describing the 1.7-GHz brightness distribution (Table~\ref{Lband-modelfit}) to the 5-GHz data, and fixed the positions of these components. We then performed a new modelfit in {\sc Difmap}, allowing only the jet component sizes, the flux densities, and, for the point-like core component, also the position to vary. The resulting difference in the position of the 1.6-GHz and 5-GHz core (0.57\,mas) is consistent with the expected value ($\la1$\,mas) as described above on the basis of the analysis by \citet{Soko11}. We applied this 0.57\,mas core shift in our reconstruction of the spectral index distribution.

As expected, the spectrum becomes gradually steeper along the jet towards the west, with the increasing distance from the core (Fig.~\ref{sp-index}). The spectral index in the core region is $-0.35$. Its sign is consistent with the value derived from the fitted flux densites of the point-like core component ($\alpha_{\rm core}$=$-0.6$). The difference in the spectral indices obtained by the two methods could be attributed to resolution effects, i.e. the blending of the core and the innermost jet component (Table~\ref{Cband-modelfit}) in the 5-GHz image restored with a more extended beam.   

The integral flux densities in our VLBI components (180.4~mJy at 1.7~GHz and 79.2~mJy at 5 GHz) are about 20--40 per cent smaller than the total flux density values obtained from the literature. This indicates the presence of jet emission on angular scales of $\sim$100~mas which is resolved out by our VLBI observations. The overall radio spectrum of J1026+2542 follows a power law with spectral index $\alpha_{\rm total}$=$-0.4$, in the observed frequency range of 151~MHz--43~GHz \citep[rest-frame frequencies from $\sim$1~GHz to 270~GHz;][and references therein]{Frey13}. Recent measurements made at 15, 31, and 91~GHz reported by \citet{Sbar13} suggest long-term variability at least at the high-frequency end in the radio, and indicate a steepening spectrum (with $-0.7$ spectral index) with increasing frequency.

\subsection{Brightness temperature limits}

With our new measurements, we are able constrain the properties of the inner radio jet in J1026+2542 in a better way than was done in \citet{Frey13}. The conclusion of that earlier study was that the VLBA data from 2006 \citep{Helm07} did not provide strong evidence for an extremely high Doppler-boosted brightness temperature in the radio jet, and thus for the small jet viewing angle found by \citet{Sbar12}. This claim \citep{Frey13} was based on the best-fitting brightness distribution model of the VLBA data with an elongated central elliptical Gaussian component.

The EVN observation at 5~GHz presented in this paper used longer interferometer baselines and therefore achieved finer angular resolution compared to the earlier VLBA experiment \citep{Helm07}. The imaging sensitivity was also improved. Moreover, component B likely moved out from the centre between the two epochs. As a consequence, the radio core (i.e. the base of the inner jet with optical depth of unity at the given frequency) can now be more reliably resolved into, and fitted with a point-like compact component and a nearby circular Gaussian (components A2 and B2 in Table~\ref{Cband-modelfit} and Fig.~\ref{VLBI-models}). Using the minimum resolvable angular size \citep{Kova05} as the upper limit for the size of A2 (major axis $\vartheta_1$=0.12~mas and minor axis $\vartheta_2$=0.09~mas; Table~\ref{Cband-modelfit}), a lower limit to its brightness temperature can be calculated as
\begin{equation}
T_{\rm B} > 1.22\times 10^{12} (1+z) \frac{S}{\vartheta_1 \vartheta_2 \nu^2} \,\, {\rm K},
\end{equation}
where $S$ is the flux density (Jy) and $\nu$ is the observing frequency (GHz). This gives $T_{\rm B}$$>$$(5.7\pm0.3) \times 10^{11}$~K for the core at 5~GHz, a lower limit exceeding that of \citet{Frey13} by a factor of six. 

Because of the steep spectrum and the larger flux density, the estimate of the brightness temperature lower limit from the 1.7-GHz EVN data is even more stringent. Considering the central point source with $S$=39.1~mJy flux density, $\vartheta_1$=0.24~mas major axis upper limit, and $\vartheta_2$=0.20~mas minor axis upper limit (Table~\ref{Lband-modelfit}), $T_{\rm B}$$>$$(22.7\pm1.2) \times 10^{11}$~K.

\section{Discussion}

\subsection{Parameters of the radio jet}

The measurement of the apparent speed in the jet ($\beta_{\rm app}$) and an estimate of the Doppler-boosting factor ($\delta$) could be applied to calculate an important physical parameter, the bulk Lorentz factor ($\Gamma$), and the viewing angle ($\theta$) of the jet with respect to the line of sight. The relevant formulae are the following \citep[see e.g.][]{Urry95,Cohe07}: 
\begin{equation} 
\Gamma=(1-\beta^2)^{-\frac{1}{2}}
\end{equation} 
\begin{equation}
\beta_{\rm app}= \frac{\beta \sin \theta}{1-\beta \cos \theta}
\end{equation}
\begin{equation}  
\delta  = \frac{1}{\Gamma(1-\beta \cos \theta)}
\end{equation}   
The method assumes an ideal (straight and narrow) relativistic beam \citep{Cohe07} in which the synchrotron-emitting plasma moves away from the central black hole with a bulk speed of $\beta$ (expressed in the units of the speed of light). In this case, the Doppler factor for the unresolved core is determined by the same physical parameters as of the more distant part of the jet. Alternatively, the jet may follow a curved (helical) trajectory, like in the case of e.g. J1159+2914 \citep{Hong04}. However, VLBI imaging data from just two epochs are not sufficient to investigate this.  

For the first time, at least tentatively, we could estimate the apparent speed of the jet components in J1026+2542. 
The jet speeds for the three components with significant values are consistent with being equal within the uncertainties (Table~\ref{properm}). For the following calculations, we adopt the median value, $\beta_{\rm app}$=12.5, as the one representative of the beam speed. In turn, the Doppler factor we can estimate from the measured core brightness temperature lower limit, $T_{\rm B} > 2.27 \times 10^{12}$~K. For this, we first have to assume a value for the intrinsic brightness temperature ($T_{\rm B,int}$). The Doppler factor is then derived as $\delta = T_{\rm B} / T_{\rm B,int}$. One possibility is to adopt the equipartition value valid for synchrotron self-absorption, $T_{\rm eq} \simeq 5 \times 10^{10}$~K \citep{Read94}. In the case of another blazar at very high redshift (J1430+4204 at $z$=4.72), a detailed study of radio jet parameters with multiple methods indicate that the equipartition value is indeed close to the true intrinsic brightness temperature \citep{Vere10}. However, other studies of large VLBI samples find it more appropriate to assume an order of magnitude higher intrinsic value, $T_{\rm B,int} = 5 \times 10^{11}$~K \citep[e.g.][]{Cohe07}, approaching the inverse-Compton limit \citep{Kell69}. We can therefore conservatively conclude that the Doppler factor is at least $\delta$$\approx$5 but could be about an order of magnitude higher.

Can these radio jet parameters be reconciled with the results found from SED modeling by \citet{Sbar13}, $\Gamma$=13 and $\theta$=3\degr? Using $\beta_{\rm app}$=12.5, the possible minimum value of the Lorentz factor is $\Gamma_{\rm min} = \sqrt{(\beta_{\rm app}^2+1)} \approx 12.5$ \citep{Urry95}. In this case, the corresponding Doppler factor is $\delta = \Gamma_{\rm min} \approx 12.5$, well within the range allowed by our brightness temperature measurements, and the viewing angle is $\theta = 4\fdg6$ \citep[$\sin \theta = \Gamma_{\rm min}^{-1}$; see Appendix A in][]{Urry95}.
We can thus conclude that the jet parameters derived from VLBI are generally consistent with those estimated by \citet{Sbar13} in a completely independent way.

By applying the method proposed by \citet{Clau13}, we give another rough estimate of the Lorentz factor. It appears from the study of the large MOJAVE sample that $\Gamma \theta_{\rm j} \approx 0.2$ for the inner jets in radio-loud AGN. Here $\theta_{\rm j}$ is the intrinsic half opening angle of the jet. For J1026+2542, we calculate it by determining the apparent half opening angle $\theta_{\rm app}=\arctan(d/2r)$ from our EVN model component diameters ($d$) and radii ($r$) at 5~GHz (Table~\ref{Cband-modelfit}). We get $\theta_{\rm app} \approx 10\degr$ after averaging over the five jet components. (This value is similar if we consider the VLBA data at 5~GHz, or our EVN data at 1.7~GHz from Table~\ref{Lband-modelfit}.) The intrinsic half opening angle is then $\theta_{\rm j}=\theta_{\rm app} \sin \theta$ \citep[e.g.][]{Push09}. Assuming $\theta=4\fdg6$ for the jet viewing angle, we obtain $\Gamma \approx 15$, in accordance with the Lorentz factor estimates cited above.   

\subsection{Apparent proper motion--redshift relation}

\begin{figure}
\centering
  \includegraphics[width=60mm, angle=270]{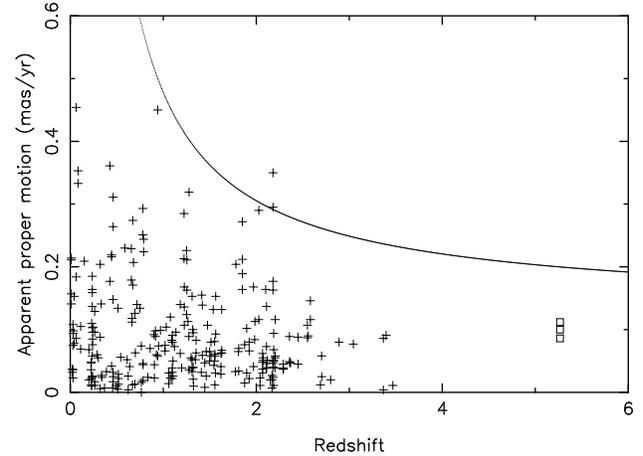}
  \caption{
Jet component proper motions measured with VLBI at 5~GHz as a function of redshift. The bulk of the data points are taken from \citet{Brit08} for comparison, showing their highest-quality (Q1) values (crosses). Our tentative proper motions derived for three components (C, D, E) in the jet of J1026+2542 are marked with open squares. The curve indicates the upper limit assuming a jet Lorentz factor $\Gamma$=25 \citep[cf. Fig.~11 in][]{Kell04} and the cosmological model adopted in this paper.
} 
  \label{proper-motion}
\end{figure}

Our measured proper motions ($\mu$$\la$0.1~mas\,yr$^{-1}$) are consistent with the trend in the apparent proper motion--redshift diagram \citep[e.g.][]{Cohe88,Kell04,Brit08} extrapolated to $z$$>$5 (Fig.~\ref{proper-motion}). In particular, if the cosmological interpretation of quasar redshifts holds, and the maximum jet Lorentz factors are not significantly higher in blazars at very high redshifts, only low values of apparent proper motions are expected there. This can be easily understood in terms of the time dilation in the expanding Universe where changes in the jet structure of a source at $z$ appear slower in the observer's frame by a factor of (1+$z$). 

There is still a long way to go until sufficient proper motion data are collected for any meaningful statistical study of the apparent proper motion--redshift relation that includes jets at very high redshifts. Multi-epoch monitoring of at least dozens of more sources would be needed to fill the gap at $3.5 \la z \la 6$ in Fig.~\ref{proper-motion}. One problem is that known radio AGN in this redshift range are scarce, especially the bright ones with rich jet structures accessible for high-resolution VLBI monitoring. Observing campaigns for obtaining spectroscopic redshifts for known VLBI sources would alleviate the situation. On the other hand, because of the time dilation, it takes at least a decade or so to get reliable proper motion estimates, even for the already known radio AGN at very high redshifts, like J1026+2542. VLBI imaging observations of a growing sample of quasars at $z>3$ started in the 1990's \citep[e.g.][]{Gurv92}. It is now time to revisit with VLBI the most distant AGN that show prominent jet structures, for measuring their apparent proper motions.

\section{Summary}

We observed the blazar J1026+2542 ($z$=5.266) at two radio frequencies (1.7 and 5~GHz) with the EVN. At 5~GHz, our observation provides the second-epoch VLBI image made after a period of 7.33~yr. Based on the data at two epochs, we detected structural changes in the jet and estimated significant apparent proper motions of three components. Such proper motion estimate was made for the first time for any AGN at $z$$>$5. The component motions are consistent with what is expected in a relativistic cosmological model. The apparent transverse speeds are superluminal, up to 14\,$c$. Compared to an earlier study \citep{Frey13}, the new 5-GHz EVN data presented here allowed us to place a more stringent constraint on the size of the compact VLBI core. The parameters of the inner radio jet are rather consistent with the Lorentz factor ($\Gamma$=13) and viewing angle ($\theta$=$3\degr$) estimated from SED fitting by \citet{Sbar13}. 
As estimated from the spectral index map, the core of J1026+2542 appears to have a radio spectrum somewhat steeper ($\alpha$=$-0.35$) than usual for blazars at lower redshifts. While this may be partly due to insufficient resolution of our data (a steep-spectrum
jet component blending with the core), the overall spectrum is clearly not dominated by a flat-spectrum core.
The turnover frequency of the system, as suggested by the total flux density measurements at multiple frequencies \citep{Frey13}, is below $\sim$1~GHz in the rest frame of the source. The comparison of the sum of the flux densities in the VLBI-detected components with the total flux densities indicates that there is significant radio emission coming from angular scales larger than $\sim$50~mas (projected linear scale $>$300~pc). This is resolved out with VLBI but could be revealed with lower-resolution interferometric imaging in the future.

\section*{Acknowledgments}

We thank the anonymous referee for helpful suggestions. The European VLBI Network is a joint facility of European, Chinese, South African, and other radio astronomy institutes funded by their national research councils. The research leading to these results has received funding from the European Commission Seventh Framework Programme (FP/2007-2013) under grant agreement No. 283393 (RadioNet3). This work was supported by the Hungarian Scientific Research Fund (OTKA, K104539).

\label{lastpage}

\end{document}